\documentclass[preprint, times]{elsarticle}

\usepackage{amsmath, amssymb}

\usepackage{graphicx}
\usepackage{dcolumn}
\usepackage{bm}
\usepackage{listings}

\journal{Physica A}

\begin{document}

\begin{frontmatter}



\title{A simple model of edit activity in Wikipedia}


\author[affi1,affi2]{Takashi Shimada\corref{cor1}}
\cortext[cor1]{Corresponding Author}
\ead{shimada@sys.t.u-tokyo.ac.jp}
\affiliation[affi1]{organization={Department of Systems Innovation, Graduate School of Engineering, The University of Tokyo}, addressline={7-3-1 Hongo}, city={Bunkyo-ku}, postcode={113-8656}, state={Tokyo}, country={Japan}}
\affiliation[affi2]{organization={Mathematics and Informatics Center, The University of Tokyo}}

\author[affi3]{Fumiko Ogushi}
\ead{ogushi@sigmath.es.osaka-u.ac.jp}
\affiliation[affi3]{organization={Center for Mathematical Modeling and Data Science, Osaka University}, addressline={1-3 Machikaneyama}, city={Toyonaka}, postcode={560-8531}, state={Osaka}, country={Japan}}

\author[affi4]{J\'anos T\"{o}r\"{o}k}
\affiliation[affi4]{organization={MTA-BME Morphodynamics Research Group, Department of Theoretical Physics, Budapest University of Technology and Economics}, addressline={Muegyetem Rkp 3}, postcode={1111}, city={Budapest}, country={Hungary}}

\author[affi5]{J\'{a}nos Kert\'{e}sz}
\affiliation[affi5]{organization={Department of Network and Data Science, Central European University}, city={Vienna}, country={Austria}}

\author[affi6]{Kimmo Kaski}
\affiliation[affi6]{organization={Aalto University, School of Science}, city={Espoo}, country={Finland}}

\begin{abstract}
A simple dynamical model of collective edit activity of Wikipedia articles and their content evolution
is introduced. Based on the recent empirical findings, each editor in the model is characterized by an ability to make content edit, i.e., improving the article by adding content and a tendency to make maintenance edit, i.e., dealing with formal aspects and maintaining the edit flow. In addition, each article is characterized by a level of maturity as compared to a potential quality needed to comprehensively cover its topic.
This model is found to reproduce 
the basic structure of the bipartite network between editors and articles of Wikipedia. Furthermore, the relation between the model parameters of editors and articles and the metrics of those calculated from the emergent network turns out to be robust, i.e. depending only on the rate of the introduction of new articles to the editing activity. This results provides us a way to relate observations in the real data to the hidden characteristics of editors and articles. For the nestedness of the networks, systems with weighted parameter distribution gives better match to the empirical one. This suggests the importance of high-dimensional nature of the ability of editors and quality of articles in the real system.
\end{abstract}




\end{frontmatter}


\section{Introduction}
Wikipedia is a paradigmatic example of a successful collective knowledge space based on ``wisdom of crowds'' principle, in which volunteer editors collaboratively edit the articles, to improve their quality in a self-organized manner~\cite{Broughton2008,Taha2013}. This complex social system of cooperative value production has attracted scientific interest
for quite some time~\cite{Mesgari2014review}. This is even more so because
Wikipedia contains the whole documentation, including all versions of the articles and the entire discussion history, is open for research~\cite{WP_download}.

One of the strengths of Wikipedia is that it is available in more than 300 languages, with
huge differences in volume and quality. A comparative analysis of mature Wikipedias reveals demographic, cultural and habitual differences in editing practice~\cite{Yasseri2012Circadian}, e.g., there is a gender imbalance in the editing activity as there are much more male editors than female ones~\cite{Wagner2015men's,Hube2017bias}. 

Considerable effort has been devoted to the understanding of the ``mystery" of the voluntary edits. Studies showed that an extreme Pareto effect is in play: A tiny minority of the editors is responsible for the majority of the edits~\cite{Priedhorsky2007Pareto}. Not only is the activity level of the editors very heterogeneous but they also perform different roles. Using machine learning categorization two large groups of activities were identified: ``Meaning preserving" (relocation, grammar, markup, and rephrase) and ``Meaning changing" (information, file, reference, wikilink, and  template)~\cite{Yang2016Who}. 

Due to the open editing model of Wikipedia one of the most relevant questions is about reliability and quality of the articles. Already early studies showed a surprisingly high level of reliability of scientific articles in Wikpedia~\cite{NatureSpecialReportOnWikipedia2005}, which has been further reinforced in later studies (see~\cite{Jemielnak2016Briding}). In Wikipedia, articles can be categorized as being ``good'' and ``featured'' based on editors' votes, such that in case of English Wikipedia somewhat more than half percent of the articles are ``good" and 0.6 thousandth is in the ``featured" category. Wikipedia has severe problems caused
by vandals and trolls~\cite{Shachaf2010Vandals} and by edit wars~\cite{Yasseri2012}, and the self-documentation of Wikipedia contains a list of controversial articles~\cite{WP_controversial} to help orientation.

While self-reported categorization is useful, an independent testing of reliability and quality of the articles is of great importance for improving the usability of Wikipedia. In any case, automated methods should be preferred because of the large number of articles and to achieve the goal of maximum objectivity. One way to go is to collect typical features of articles, like ``templates"~\cite{Wong2021reliability} or popularity~\cite{Lewoniewski2020Reliability} and to use machine learning techniques for the categorization. The assessment of the quality of articles is an even more demanding task. Machine learning techniques have been applied to solve this problem as well~\cite{Adler2008quality,Lipka2010,Cosley2013quality,Dang2016quality}. While some of these methods produce remarkable results in terms of precision, the nature of the approach does not allow for an insight into the mechanism leading to better or worse articles. 
It was shown that cooperation and quality are 
closely related~\cite{Wilkinson2007quality}, 
therefore,
to model and understand the process it is necessary to consider the complex social dynamics of value creation. For the study of controversial articles this approach has already proved to be fruitful~\cite{Yasseri2012,JTorok2013,Gandica2014}.

Recently we showed that the editor-article bipartite network based on edit records enables to introduce simultaneous measures for both editors and articles that 
can be calculated in a self-consistent way~\cite{Ogushi2021}. In that work, we proposed that editors at times edit an article to enrich its content, while at other times they make edits to maintain the format of an article.
In the following we call the former type of edit as ``content edit'' and the latter type of edit as ``maintenance edit''.
Based on this idea, we introduced a self-consistent metrics for evaluating editor's higher tendency to make a maintenance edit, {\it scatteredness}, and the resulting quality 
of an article, {\it complexity}.
With the proposed self-consistent measure one can find articles, which are not prominent in the number of edits or page views but contain complicated and/or professional information. Therefore, the  {\it complexity} can differentiate the goodness of an article from its mere popularity or 
controversiality. Furthermore, the {\it scatteredness} provides a way to characterize an editor's edit pattern and activity.

The main aim of this study is to introduce a simple model, which could shed light or explain the character of the above observed Wikipedia edit patterns.
For this, our model should provide the following qualitative properties, which bridges the theoretical (hidden) characters of editors and articles and the characteristics in the resulting edit patterns.
First, the basic network characteristics such as the degree and strength distributions of editors and articles should be reproduced. And in that emergent network, the editor's higher tendency to make maintenance edits and the scatteredness measured from the network should be positively correlated. At the same time, the editor's higher ability/expertise and the scatteredness should be negatively correlated. For articles, the intrinsic character of its quality or complexity should be positively correlated with the complexity measured from the edit records.

This paper is organized such that in the next section 2, we describe the self-consistent metrics that 
was used for the empirical analysis of Wikipedia edit network. Next in section 3, we introduce our simple agent-based model of Wikipedia edit process. This is followed in section 4 with the analysis of simulation results to show that our model is capable of reproducing the edit pattern, and relating that to the character of editors and articles. Finally in section 5 we draw conclusion and present discussions.

\section{Complexity-Scatteredness Measure}
In order to evaluate the tendency of an editor to make a maintenance edit, which we will call {\it scatterdness}, we calculate the inverse sum of the {\it complexity} of the articles the editor has edited. This is because the set of articles an editor has edited should be more in number and more scattered among articles with broader {\it complexity} than in case of making mainly content edits on selective articles according to editor's field of expertise. The {\it complexity} of each article is estimated from the sum of the inverse of {\it scatteredness} of the editors who have edited that article, i.e. the expected net contribution of content edits the article has received.
More specifically, the self-consistent metrics are defined by the links of a bipartite edit network between $N_E$ editors and $N_A$ articles as follows
\begin{equation}
    \tilde{D}_e^{(n)}
    =
    \sum_{\alpha}^{N_A} \frac{\displaystyle w_{e \alpha}}{\displaystyle C_\alpha^{(n-1)}},
	\quad
	\tilde{\mbox{$C$}}_\alpha^{(n)}
	= \sum_{e}^{N_E} \frac{\displaystyle w_{e \alpha}}{\displaystyle D_e^{(n-1)}},
    \label{eqs_def_SC}
\end{equation}
where $w_{e \alpha}$ is the number of edits an editor $e$ made on an article $\alpha$. $D_e^{(n)}$ and $C_\alpha^{(n)}$ are the {\it scatteredness} of the editor $e$ and the {\it complexity} of the article $\alpha$ respectively, calculated after $n$ times of recursive evaluation steps.
Note that after each re-evaluation step, we normalize the scatteredness and the complexity measures in Eq. (\ref{eqs_def_SC}) to read as follows
\begin{equation}
	D^{(n)}_e
    =
	\frac{\tilde{D}^{(n)}_e}{\displaystyle \frac{1}{N_E} \sum_i^{N_E} \tilde{D}^{(n)}_i},
	\quad
	C^{(n)}_\alpha =
	\frac{\displaystyle \tilde {C}^{(n)}_\alpha}{\displaystyle \frac{1}{N_A} \sum_\xi^{N_A} \tilde{C}^{(n)}_\xi}.
	\label{eqs_normalization_SC}
\end{equation}
Starting from the uniform initial condition, $D^{(0)}_e = 1, \ C^{(0)}_\alpha = 1$, the proposed recursive process on empirical data yields good convergence for both the scatteredness and complexity, as discussed in~\cite{Ogushi2021}. Although the scatteredness of editors and the complexity of articles are correlated with their number of edits (i.e. strength) as is expected from the definition, they show considerable variations, which were 
found to give rich information about the editors and articles. For example, one can differentiate human-labeled ``featured'' articles from high-complexity non-labeled articles. Furthermore, among ``controvertial'' articles, those that have higher complexity relative to their number of edits tend to have scientific content. 

\section{Model}
\subsection{Editors and Articles}
We model the state of each article with a single dynamical variable, {\it quality}, which indicates the amount and/or quality of information the article contains.
The {\it quality} of an article $\alpha$, which will be denoted as $q_\alpha$, starts from $0$ at the moment of its creation
and it will be increased by the contents edits, up to its intrinsic maximal potential {\it quality} $Q_\alpha \in (0, 1)$.
The idea behind the varying maximal quality is that there are subjects for which one does not need much content or professional expertise to complete the edit, e.g. the page of a small unimportant village.
It should be noted that neither 
the potential nor the present quality of an article can be determined directly from Wikipedia data.
However, the human-aided labeling of articles as ``featured'' and ``good'', can be used as an incomplete or partial measure for the latter.

In the model, each editor is characterised by two parameters. The first one is the content edit ability $A_e \in (0, 1)$ that denotes the editor $e$'s 
expertise. The editor can edit the article only if its ability is larger than the present state of the article, i.e. $A_e > q_\alpha$.
The second parameter of an editor is $M_e$, which characterizes the editor's tendency to make maintenance-type edit. The maintenance edit of the article does not improve its quality and the editors can perform this type of activity on subjects beyond their expertise (e.g. perfecting grammar, unifying the notation or styling, etc.).

In reality the {\it quality} of an article and the {\it ability} of an editor should be considered multi-dimensional in nature where the dimensions represent the different fields of human knowledge. However, for simplicity, we here consider them to be single dimensional.

\subsection{Dynamics of the Model}
Let us consider initially a small network consisting of $N_{E} = 1,000$ editors and $N_{A}^{ini} = 10$ articles. In this initial network, each article has only one link from a randomly chosen editor, regarded as the initial creator of that article.

In the dynamics 
at each time step, the following edit activity process is performed for $T$ times: A randomly chosen editor tries to make a content edit on some selected set of articles provided that its edit ability $A_e$ exceeds the current quality of the articles, after which the editor may perform a number of maintenance edit operations according to the editor's tendency $M_e$.
After each time step of $T$ single-editor edit activities, a new article is added to the system. Then 
an initial link is assigned at random 
with a probability proportional to the editors' total number of edits, which corresponds to assuming that the editor's tendency 
in creating a new article is proportional to the editor's past edit activity.

This 
network evolution process is repeated until the number of articles reaches $N_A \gg N_A^{ini}$. Therefore, in the end of the simulation, we obtain a bipartite network between $N_E$ editors and $N_A$ articles, with the average strength of editors $\langle s_E \rangle \approx \left( \frac{N_A}{N_E} \right) T$ and articles $\langle s_A \rangle \approx T$, respectively.

\begin{figure}[tbh]
\begin{center}
\includegraphics[width=0.9\linewidth]{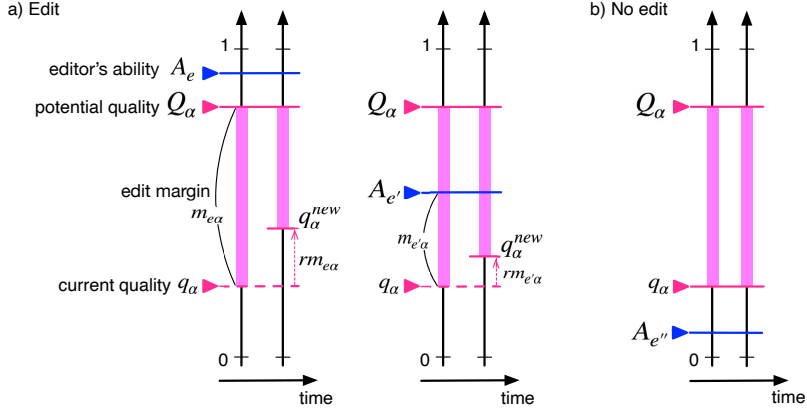}
\caption{Schematic image of the contents edit dynamics. (a) The editor can make the contents edit on articles whose current quality $q_\alpha$ is lower than the editor's ability $A_e$. The increment of the quality by the edit is proportional to the margin left.
(b) The editor cannot make contents edit if the article's current quality is higher than the editor's ability.}
\label{fig_editmodel}
\end{center}
\end{figure}

\subsection{Edit Process}
In the bipartite network of editors and articles, the edit process of each editor consists of two steps, namely first making an attempt to do a content edit
followed by a maintenance edit. Next we describe these two processes separately.
\subsubsection*{Content Edit}
The randomly chosen editor $e$ looks at $n_s$ randomly chosen articles, to find the one for content edit in that round. For this, the editor calculates the margin for edit on each article $\alpha$,
\begin{equation}
m_{e \alpha} = \min\left( A_e, \ Q_\alpha \right) - q_\alpha,
\end{equation}
as depicted in Fig.~\ref{fig_editmodel}~(a).
If the margin is non-positive, there is no room left for the editor $e$ to contribute by a content edit to the article $\alpha$. From the articles yielding positive margins the editor chooses the article $\alpha_*$ that gives maximum edit margin,
\begin{equation}
m_{e\alpha_*} = \max\left( \{ m_{e\alpha} \}_{n_s}\right),
\end{equation}
and performs a content edit on it. By the content edit, the quality of that article $\alpha_*$ is increased by a fixed proportion $r$ of the margin, i.e.
\begin{equation}
q_{\alpha_*}^{new} = q_{\alpha_*} + r m_{e \alpha_*} \ \left( 0 < r \le 1 \right).
\end{equation}
If an editor cannot find any editable one among the randomly searched $n_s$ articles, no content edit is performed at that round (Fig.~\ref{fig_editmodel}~(b)).

In what follows the parameters $n_s$ and $r$ are assumed to be uniform among the editors.
Because the average number of edits per article evolving under this dynamics is roughly proportional to the edit activity times $T$, its product with the proportion of the edit, $r T$, characterises the average maturity of article, i.e. on average how near the article are to the state of its ideal completion.

\subsubsection*{Maintenance Edit}
After performing a trial of content edit, each editor may also make a maintenance edit with a probability $P_M$ calculated based on the editor's intrinsic tendency and the effort already expended during the content edit, as follows
\begin{equation}
    P_M = M_e - f_e,
\end{equation}
where
\begin{equation}
    f_e =
    \begin{cases}
     0 & \mbox{no content edit}\\
     m_{e\alpha_*} & \mbox{after content edit}
     \end{cases}.
\end{equation}

The article on which the maintenance edit is performed, is selected with probability proportional to its strength. The motivation behind this choice is the fact that more editing generally requires more tidying up and also that popular articles are visited more often also by the editors who can spot inaccuracies.

Note that the quality of the article $q_\alpha$ does not change by making a maintenance edit. Once an editor embarks on doing maintenance edit it may continue with this activity \cite{karsai2012universal}. To realize this one can consider 
the effort an editor needs for a maintenance edit, which 
can be modelled by assuming $f_e$ being incremented with $\xi$.
As the effort needed to make maintenance edit is considerably smaller than that of the content edit,
we will take $\xi = 0$ for simplicity.
Therefore the editor may make another maintenance edit with the same probability $P_M$, otherwise this edit activity process ends. Note that the results in the following section do not change for non-zero but very small $\xi \ll 1.$

\section{Results}
Let us first examine the case in which all $A_e, M_e,$ and $Q_\alpha$ are uniformly distributed in the range $(0, 1)$.
The system size is set to be $N_E = 1,000$ and $N_A = 1,270,000$ for making direct comparison with the previous empirical study of English Wikipedia.
Other parameters are chosen to be $N^{ini}_A = 10, n_s = 10, r = 0.02,$ and $T= 20$, unless otherwise noted.

\subsection{Basic Structure of the Emergent Network}
\subsubsection*{Strength Distributions}
\begin{figure}[bth]
\includegraphics[width=0.49\linewidth]{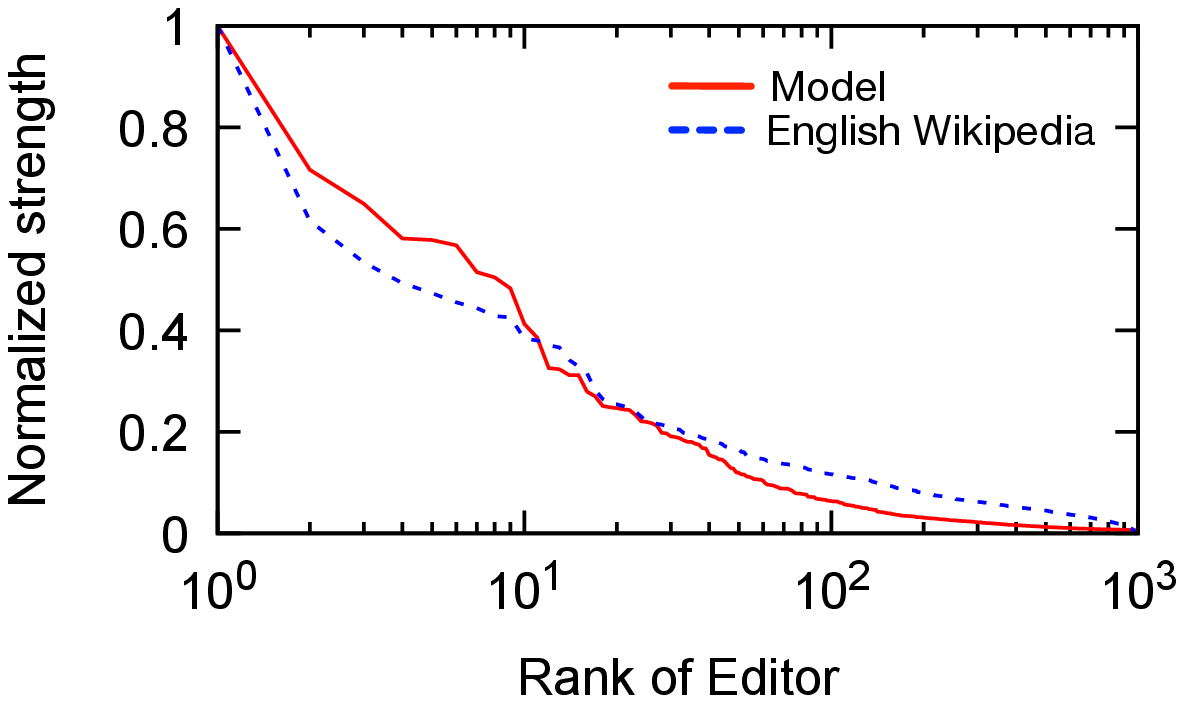}
\includegraphics[width=0.49\linewidth]{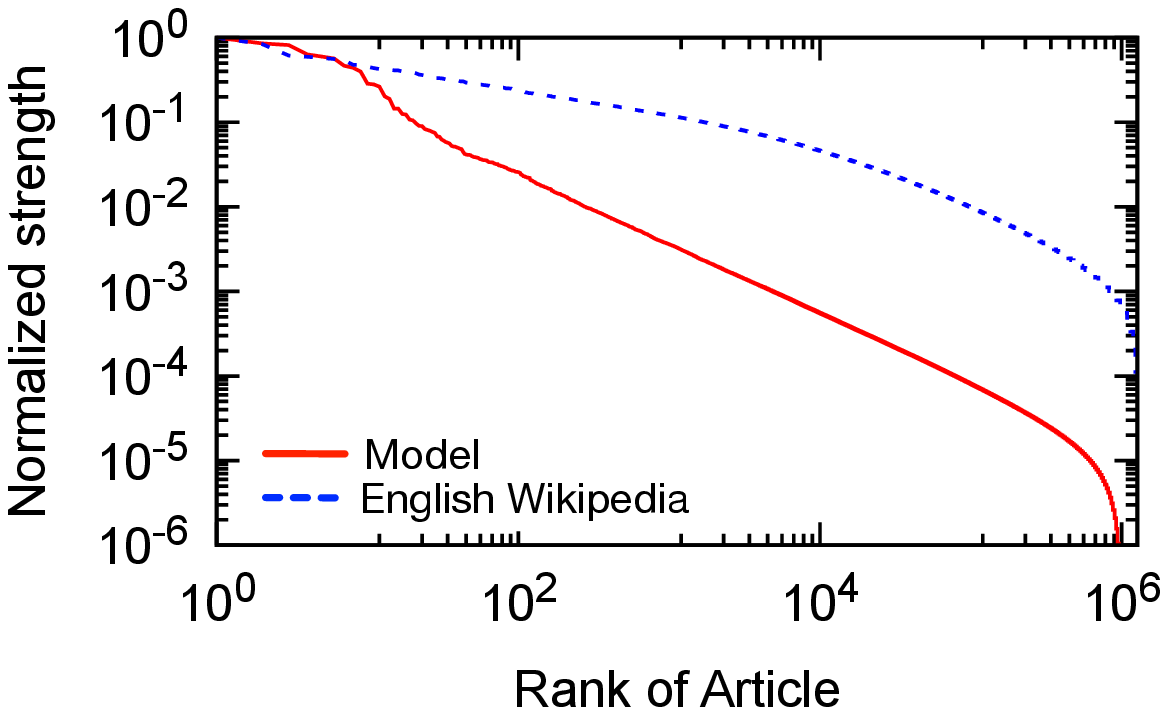}
\caption{Comparison of the strength distributions of the (left) editors and (right) articles of the model with the empirical data.}
\label{fig_strength}
\end{figure}

In Fig.~\ref{fig_strength} we show the normalised strength distributions (i.e. distributions of the number of edits) vs. the rank of editors and articles obtained from the present model and compared with the analysis of English Wikipedia. The distribution in the editor side is found to be similar to that of empirical Wikipedia network for $N_E = 984$ top editors and about $1.27 \times 10^{6}$ articles.
The distribution in the article side is also similar for the bottom 99\% of articles (with rank $>10^4$) in the power-law shape and its exponent around $-1$. This power-law in the model stems from the linear dependence of the maintenance edit activity on the present number of edits of the article (i.e. Yule-Simon process ~\cite{Hashimoto2016PRE}).
The agreement with the empirical data implies similar linear edit activity dependence in it, though in this case we cannot clearly distinguish maintenance and content edits. On the other hand for the
top 1\% articles, empirical data show slower power-law decay with the exponent $\sim -1/3$.
The regime in which similar slow decay is obtained in the model corresponds to the initial number of articles ($N_A^{ini}$), which implies the effect of the initial foundation period of Wikipedia. 

\subsubsection*{Scatteredness and Complexity}
\begin{figure*}[bth]
\begin{center}
\includegraphics[width=0.9\linewidth]{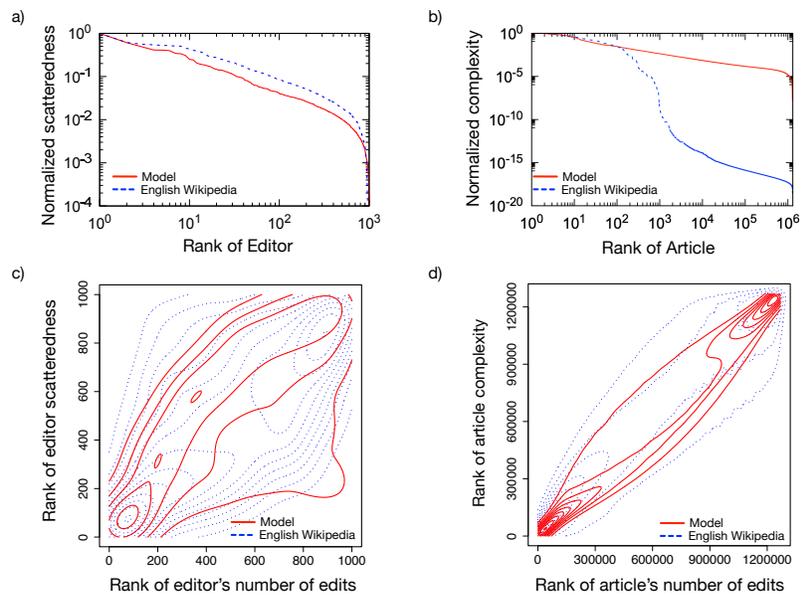}
\caption{The distributions of the editors' scatteredness and the articles' complexity. Rank plots of (a) the editor's scatteredness and (b) the article complexity. Density distributions of (c) editors and (d) articles, in the relative rank-rank plane of its number of edits and the metrics.
}
\label{fig_SC}
\end{center}
\end{figure*}

The distributions of the normalised scatteredness and complexity calculated from the emergent network vs. the rank of editors and articles compared with those from the analysis of English Wikipedia are depicted in Fig.~\ref{fig_SC}.
The scatteredness of the editors shows surprisingly good agreement with the empirical data (Fig.~\ref{fig_SC} (a)). On the other hand, the complexity calculated from the model network shows good agreement for the top 100 ranked empirical articles, after which the empirical complexity shows a dramatic drop. The agreement in the slope is broken till the rank of article reaches around 1,000 (Fig.~\ref{fig_SC} (b)).
This difference is considered to be originating from the nestedness of the article side, which we will treat later.
Although the scatteredness and complexity in the model network are correlated with the strength, as expected from the definition (Eqs. (\ref{eqs_def_SC})), they show considerable variation both for the model and empirical networks (Fig.~\ref{fig_SC} (c), (d)). This variation is a necessary condition for the metrics to provide more information about the editors and articles, the significance of which will be confirmed in the following sections.

\clearpage
\subsection{Relation between Model Parameteres and Empirical Metrics}
\subsubsection*{Editor Parameters and Scatteredness}
\begin{figure}[bth]
\begin{center}
\includegraphics[width=0.45\linewidth]{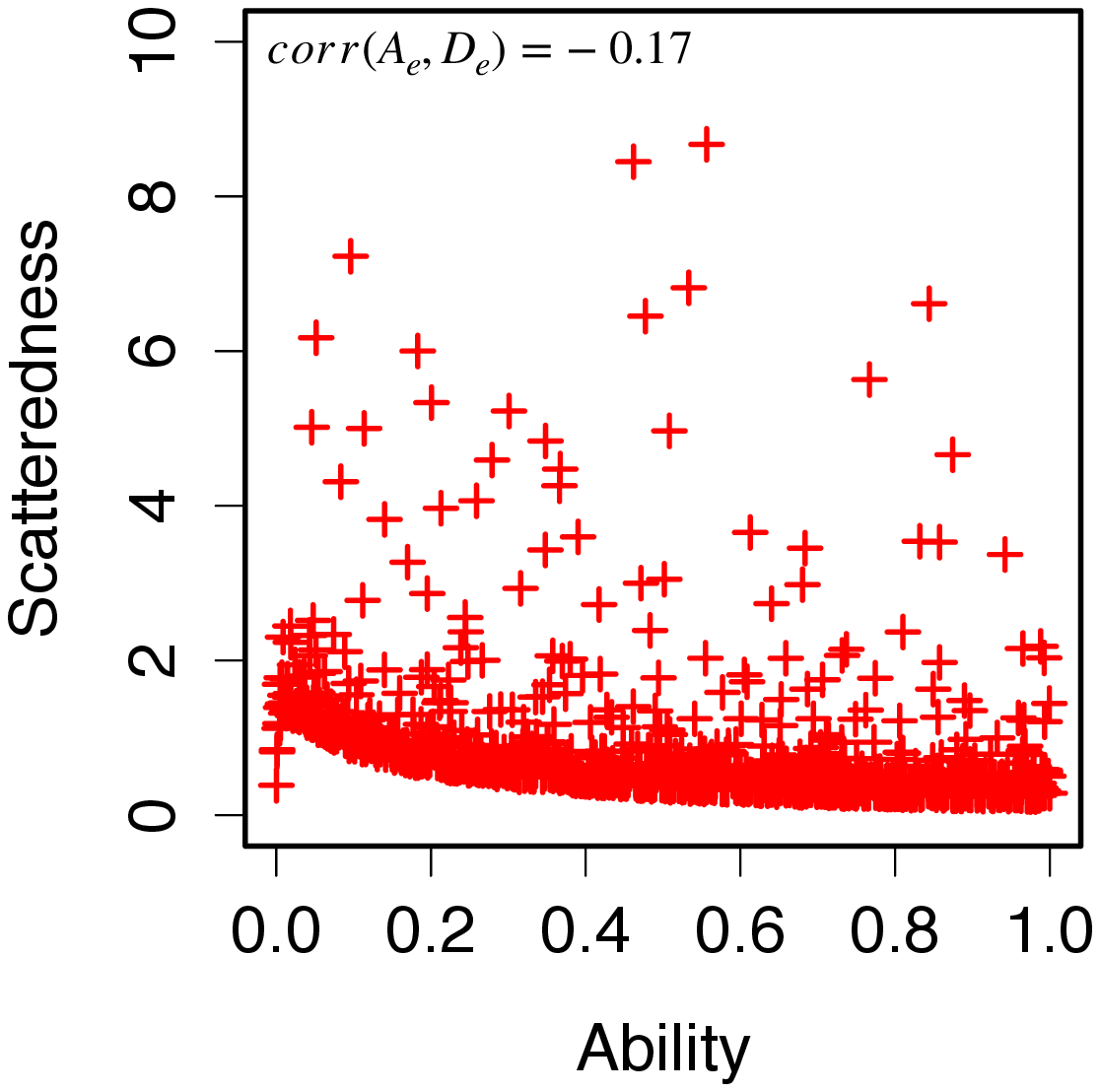}
\includegraphics[width=0.45\linewidth]{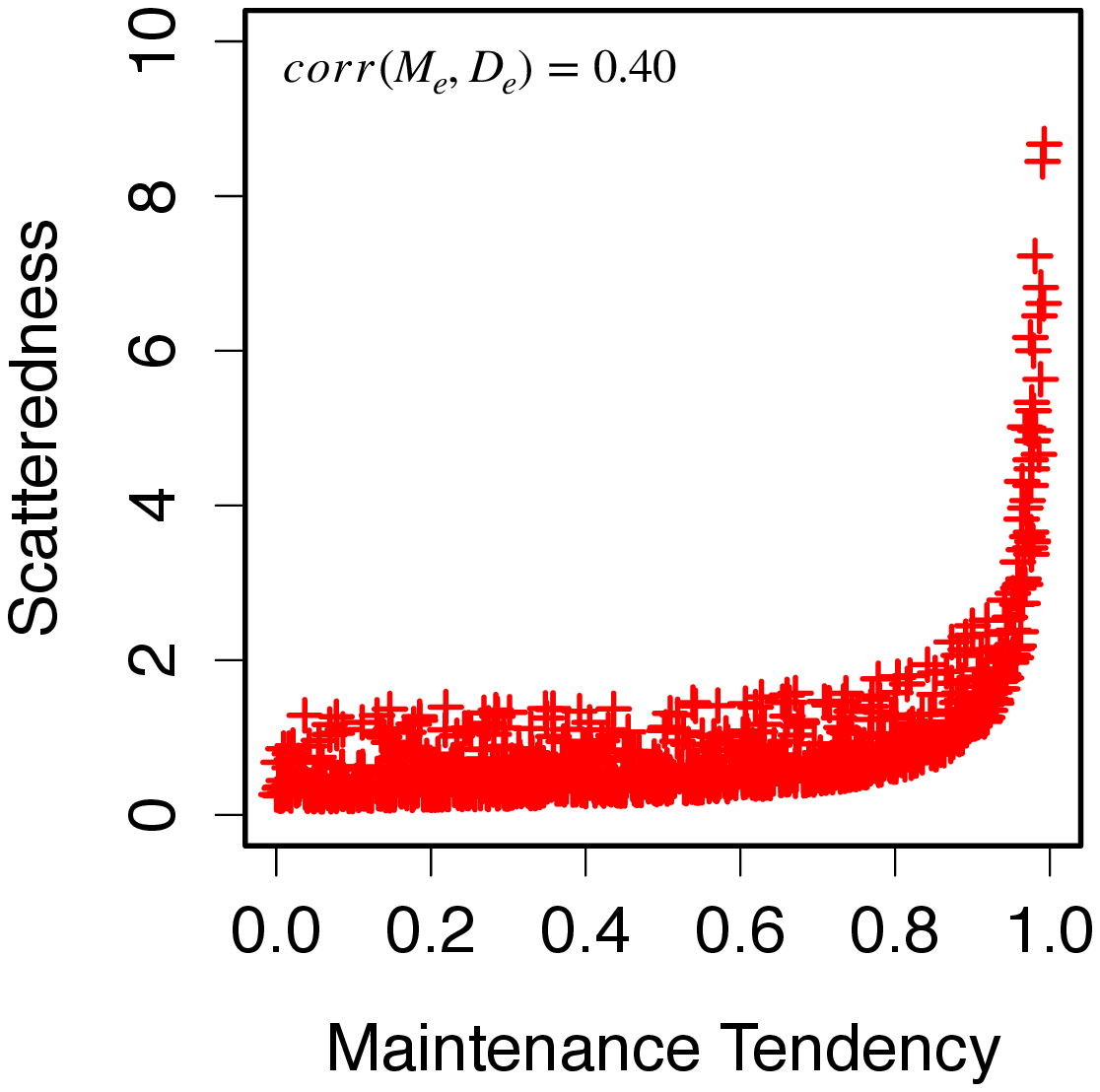}
\caption{Editor's scatteredness $D_{e}$, to its edit ability $A_e$ (top) and to maintenance tendency $M_e$ (bottom) for the system with $N_E = 1,000, N_A = 1,270,000, n_s = 10, r = 0.02,$ and $T = 20$. The editors' scatteredness $D_{e}$ is negatively correlated with $A_{e}$ (
$corr(A_{e}, D_e) = -0.17$)
and positively correlated with $M_{e}$ ($corr(M_{e}, D_{e}) = 0.40$).}
\label{fig_scatt}
\end{center}
\end{figure}

In Fig.~\ref{fig_scatt} we show how the editors' model parameters, i.e. the edit ability $A_e$ and maintenance tendency $M_e$, are reflected in their scatteredness measured from the emergent bipartite network. 
The figures show that the editors' scatteredness is negatively correlated with the editors' ability but positively correlated with their maintenance tendency.
Hence we can regard the higher scatteredness of a model editor as a signature of lower ability in making content edits and/or higher tendency of making maintenance edits.

\subsubsection*{Article parameters and the metrics}
\begin{figure*}[bth]
\begin{center}
\includegraphics[width=0.8\linewidth]{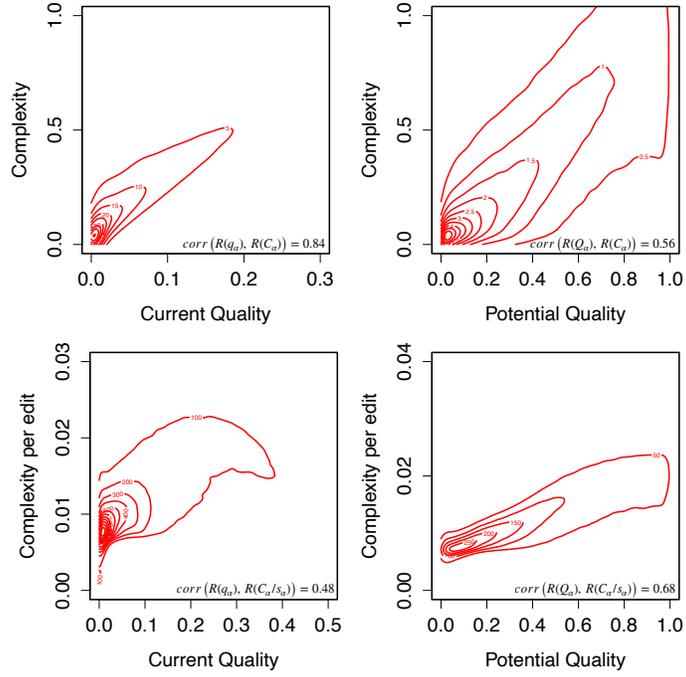}
\caption{Article complexity $C_\alpha$ measured from the model network to its current quality $q_\alpha$ (top left) and to potential quality $Q_\alpha$ (top right). Complexity normalized by its number of edits (strength), $C_{\alpha}/s_{\alpha}$, are also plotted in bottom panels. Article complexity $C_{\alpha}$ and normalized complexity $C_{\alpha}/s_{\alpha}$ are positively correlated with article quality.
}
\label{fig_complexity}
\end{center}
\end{figure*}

The article complexity in our model is found to be positively correlated with its quality $q_\alpha$ and the potential quality $Q_\alpha$, as it was postulated in the previous empirical study (Fig.~\ref{fig_complexity}).
The correlation with the article quality is larger for the complexity normalized by strength, $C_\alpha / s_\alpha$. This is also consistent with the reported fact that the non-parametric measure, defined as the rank ratio of complexity and strength of articles $J_\alpha = R^C_\alpha /R^s_\alpha$, can be used to find the high quality or speciality articles.

\subsection{The Effect of the Growth of the System}
Next we examine the dependence of these results, namely the correlations between the parameters ($A_e, M_e, q_\alpha$) and the observed scatteredness and complexity, on the other model parameters. While $N_E, N_A,$ and $n_s$ are found to be irrelevant in the regime where $N_E \gg 1, N_A \gg 1,$ and $n_s > 2$, the parameter $rT$ that characterizes the average maturity of articles (i.e., how near it is to its potential maximum quality), is found to take the role of the control parameter. As described in the Model section, if $rT\ll1$ then new articles appear before the existing ones were exclusively edited. We call this regime the non-matured, and the opposite case mature.

\begin{figure*}[tbhp]
\includegraphics[width=1.0\linewidth]{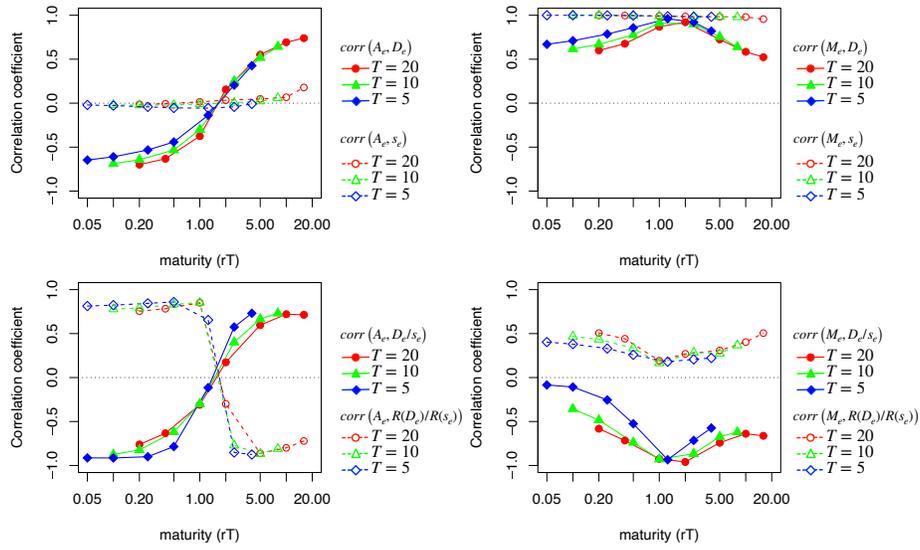}
\caption{The dependence of the correlation between editor's characteristics ($A_{e}$ and $M_{e}$) and those metrics ($D_{e}$, $s_{e}$, $D_{e}/s_{e}$, and $R(D_{e})/R(s_{e})$) on the maturity parameter $rT$.
Other system parameters are set to be $N_E = 1,000, N_A = 1,270,000$, and $n_s = 10$.}
\label{fig_rT_editor}
\end{figure*}

\begin{figure*}[tbhp]
\includegraphics[width=1.0\linewidth]{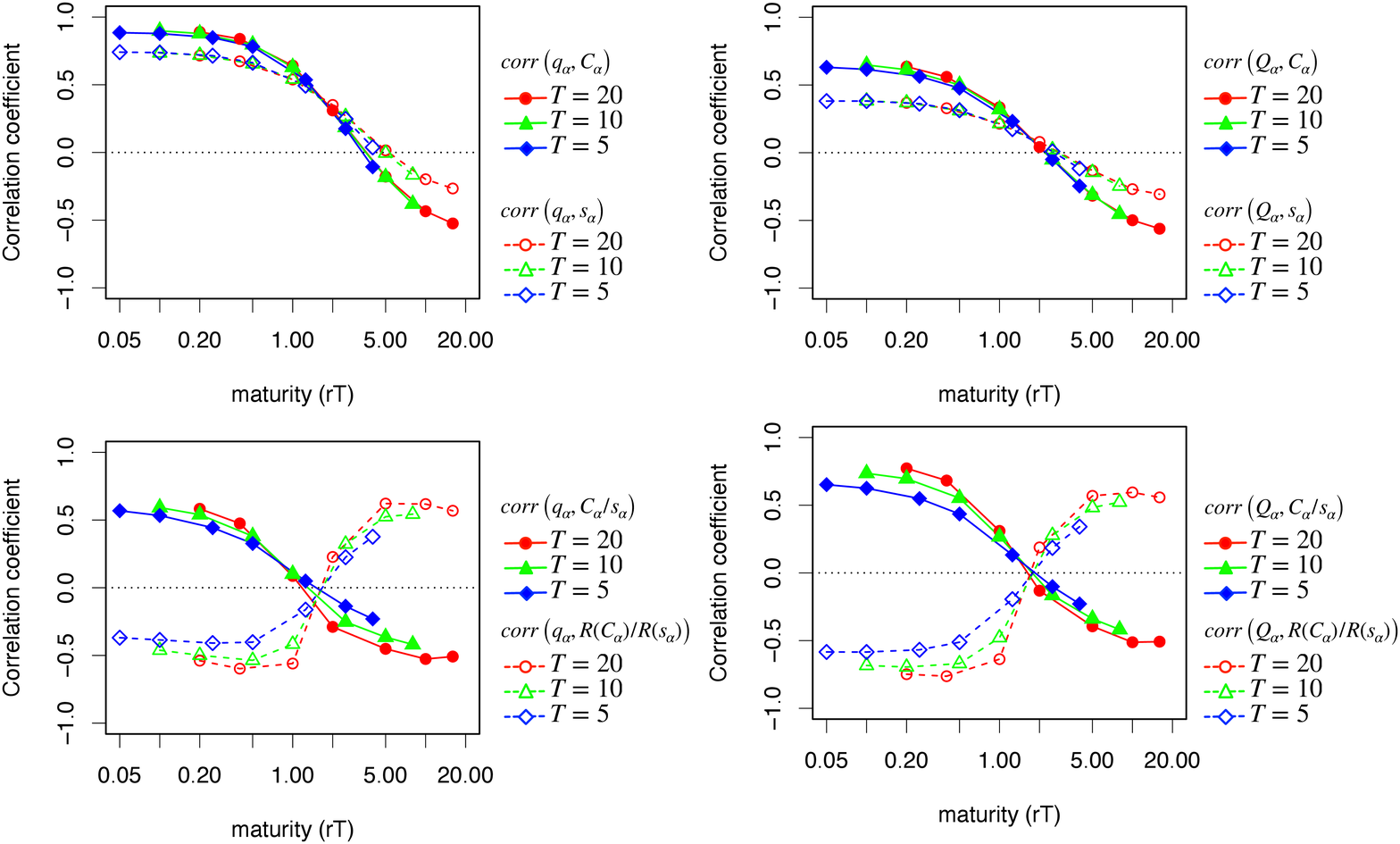}
\caption{The dependence of the correlations between article's quality and the metrics of the article ($C_{\alpha}$, $s_{\alpha}$, $C_{\alpha}/s_{\alpha}$, $R(C_{\alpha})/R(s_{\alpha})$) on the maturity parameter $rT$. Other system parameters are set to be $N_E = 1,000, N_A = 1,270,000$, and $n_s = 10$.}
\label{fig_rT_article}
\end{figure*}

In Figs~\ref{fig_rT_editor} and ~\ref{fig_rT_article} we show the Pearson correlation coefficient for different measure pairs. We can observe that most of the correlations change 
signs 
at $rT \simeq 1$ to $2$.
For example the scatteredness and editor's edit ability are anti-correlated in the non-mature regime, which is the more natural one where editors with high ability tend to select 
articles with high potential quality for content edit. 
On the other hand editors with limited knowledge tend to search 
around for possibilities to make content edit. 
In the mature regime 
the experienced editors also have to look for an article, in which 
improvement is still possible thus scattering their work.
The above correlation ensures that in the non-mature regime the article quality and complexity are correlated, as high quality articles are mainly edited by able editors which have low scatterdness. This then reversed to an anti-correlation in the mature regime.

We can conclude that our model has different phases with often opposite correlation between measures and editor behavior, with the transition point around $rT \simeq 1$ to $2$.
We could fit the empirical Wikipedia data by choosing $rT=0.4$ from which we can conclude that Wikipedia is in the non-mature state. It would also be interesting to study other cooperative value production systems, e.g. GitHub, that could be in the mature state.

%
%


\subsection{Toward a High-dimensional Model}
So far we have seen that the present simple model of the editors' edit dynamics can reproduce the basic structure of Wikipedia edit network. In this the editor's edit ability and the inverse scatteredness, editor's maintenance edit tendency and the scatteredness, and the quality of articles (both the current one and its potential maximum) and the complexity are all positively correlated.
This relation holds as long as the system parameters are in the ``growth phase'' ($rT < 1$), meaning that most articles are not near its potential completion due to the successive creation of new articles.

\begin{figure}[tbhp]
\begin{center}
\includegraphics[width=0.7\linewidth]{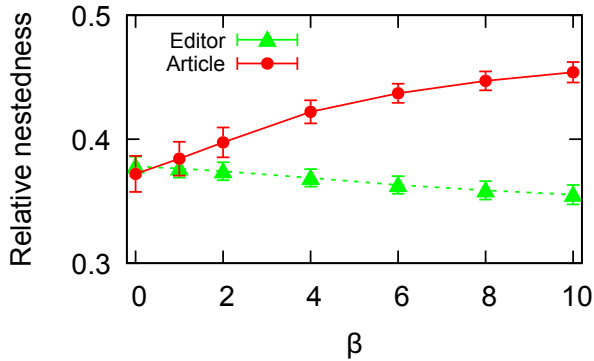}
\caption{Relative nestedness versus the exponent $\beta$ for the distribution of article's quality. Other system parameters are set to be $N_E = 1,000, N_A = 1,270,000$, and $n_s = 10$.
}
\label{fig_nestedness_to_beta}
\end{center}
\end{figure}

One characteristic left to be explained is the nestedness of the network.
The average nestedness for editors and articles in English Wikipedia, relative to the ones of configuration model with the same strength distributions, are $\tilde{\eta}^E_{WP} = 0.75$ and $\tilde{\eta}^A_{WP} = 1.60$, respectively.
Comparing to that, those values of the model network,
$\tilde{\eta}^E_{model} \sim \tilde{\eta}^A_{model} \sim 0.4$,
turn out to be smaller and indistinguishable from 
each other. One of the possible reasons for this difference could be the simplicity of the model to consider the editor's ability space and the article's quality space being one dimensional. To test this, we investigate the case that the potential quality of articles obeys the Boltzmann distribution,
\begin{equation}
    P(Q_\alpha)
    = \left[ \frac{\beta }{1 - {\rm e}^{-\beta}} \right] {\rm e}^{-\beta Q_\alpha}
    \quad
    \left( P(> Q_\alpha) = \frac{1 - {\rm e}^{-\beta x}}{1 - {\rm e}^{-\beta}} \right),
\end{equation}
meaning that there are more easy (low potential quality) articles than the articles that 
need high ability to be completed. Here the inverse-temperature $\beta \ge 0$ controls the amount of unevenness. This type of weight in distribution should be also true if the requirement for the expertise to make the content edit consists of multiple dimensions (i.e. expertise in multiple different fields such as physics, chemistry, history, etc.).
As shown in Fig.~\ref{fig_nestedness_to_beta}, the nestedness for the systems with $\beta > 1$ becomes larger in the article side than in the editor side, which is qualitatively consistent with the analysis results in the case of the empirical data.

\section{Conclusions}
We have introduced a simple model of the editing dynamics of the Wikipedia articles. Despite its simplicity having only one parameter for an article and two for an editor, the model reproduces the basic structure of the bipartite network observed in the case of English Wikipedia.
In addition the relations between the model parameters and the empirical metrics gives a plausible picture to understand why the complexity-scatteredness metrics were successful in characterizing the editors and articles in empirical data, such as in finding professional articles~\cite{Ogushi2021}.

Our model also shows the importance of growing or evolving 
nature of the English Wikipedia, which leaves room for the completion of edits for most articles.
In the corresponding regime of the model, the growth phase ($rT < 1$), an editor's higher ability for a content edit and lower tendency 
for a maintenance edit is reflected as lower scatteredness, and an article's higher quality tends to result in its higher complexity.
However, this relation becomes opposite in the regime where the edits on all the articles are almost completed ($rT > 1$).

Our model still leaves some unsolved points that could with further analysis enhance understanding of the edit dynamics. 
Especially, in the case of an empirical network, the nestedness in the article side is larger than that in the editor side, while the ones in the model are either even or 
smaller. This difference is also regarded as the reason for the disagreement found in the width of complexity distribution (Fig. \ref{fig_SC} (b)).
This is because the rewired empirical network, in which the nested structure is broken while the strength distribution is kept fixed, gives rise to the complexity distribution similar to that of the model~\cite{Ogushi2023}.
In the model with weighted distributions for $Q_\alpha$, with 
the low-quality topics being 
more densely populated, we obtain better results for the nestedness in the article side.
This implies the importance of higher dimension for the quality of an article and hence for the corresponding ability required from an editor to make a content edit, to describe the multiple different fields of knowledge, taxonomic structure of subjects, etc.

The 
present model does not have enough features for direct comparison with the real-time dynamics~\cite{Yun2016}.
The model also does not include any mechanism for explaining the correlated bursty edits among some editors on an article that became controversial~\cite{JTorok2013}. Testing the effect of such dynamical mechanism would be an interesting future work.

These findings should be tested by more detailed information of edit activity in the empirical data.
It also serves as a good basis for extending the model towards a realistic one with more complex editor and article characteristics.

\section*{Acknowledement}
TS and FO thanks to the support by JSPS KAKENHI grant number 21K19826. FO was supported by JST, PRESTO grant number JPMJPR2121, JAPAN. JK acknowledges support from project EU H2020 Humane AI-net (Grant No. 952026), and from Horizon 2020 under grant agreement ERC No 810115 - DYNASNET. JT thanks the support of the Ministry of Culture and Innovation and the National Research,
Development and Innovation Office under Grant Nr. TKP2021-NVA-02.




\bibliographystyle{elsarticle-num} 
\bibliography{ref.bib}


\end{document}